\documentclass[showpacs,twocolumn]{revtex4}
\usepackage{graphicx}
\usepackage{amsmath}
\usepackage{amssymb}
\usepackage{epsfig}

\newcommand{\be}{\begin{equation}}
\newcommand{\ee}{\end{equation}}
\newcommand{\ba}{\begin{eqnarray}}
\newcommand{\ea}{\end{eqnarray}}

\newcommand{\mev    }{\ensuremath{\mathrm{MeV}}}

\newcommand{\gev    }{\ensuremath{\mathrm{GeV}}}

\newcommand{\evsq  }{\ensuremath{\mathrm{eV^2}}}

\newcommand{\fb     }{\ensuremath{\mathrm{fb}}}
\newcommand{\der}{\ensuremath{{\operatorname{d}}}}

\newcommand{\numu}{\ensuremath{\nu_\mu}}

\newcommand{\fSK}{\ensuremath{f_{\rm SK}}}
\newcommand{\ormu}{\ensuremath{1\rm{R}\mu}}
\newcommand{\mmu}{\ensuremath{m_{\mu}}}

\begin{document}
\title{Muon spectra of Quasi-Elastic and 1-pion production events 
at the KEK LBL neutrino oscillation experiment}

\author{
Ji--Young\ Yu$^1$, E.\ A.\ Paschos$^1$, D.\ P.\ Roy$^2$, 
I. Schienbein$^3$
}
\vspace{1cm}
\affiliation{$^1$ Theoretische Physik III, University of Dortmund, 
44221 Dortmund, Germany
\\
$^2$ Tata Institute of Fundamental Research, Bombay-400005, India
\\
$^3$ DESY/University of Hamburg, Notkestrasse 85, 
22603 Hamburg, Germany
\\
Presented by Ji-Young Yu at ICFP03, October, 2003, Seoul/Korea
}

\begin{abstract}
\noindent 
We present predictions for the flux averaged muon energy spectra 
of quasi-elastic (QE) and 1-pion production events for the 
K2K long-baseline experiment.
Using the general kinematical considerations 
we show that
the muon energy spectra closely follow
the neutrino energy spectrum with downward shift of the energy
scale by $0.15\ \gev$ (QE) and $0.4\ \gev$ (1-pion production).
These predictions seem to agree with the observed muon 
energy spectra in the K2K nearby detector.
We also show the spectral distortion of these muon energy spectra 
due to the neutrino oscillation for the SK detector.
Comparison of the predicted spectral
distortions with the observed muon spectra of the 1-Ring and
2-Ring muon events in the SK detector will help to determine the 
oscillation parameters. 
The results will be
applicable to other LBL experiments as well.   

\end{abstract}
\pacs{13.15.+g; 25.30.Pt; 95.55.Vj}
\maketitle


\section{Introduction}

%

Recently the KEK to Kamioka long-baseline neutrino
oscillation experiment (K2K) has published its first result
\cite{Ahn:2002up}, which confirms the existence
of $\numu$ oscillation as seen in the Super-Kamiokande
(SK) atmospheric neutrino data \cite{Fukuda:1998mi}.
The observed oscillation parameters from K2K agree well 
with the neutrino mass and mixing
angles deduced from the atmospheric neutrino 
oscillation data \cite{Fukuda:1998mi}
\begin{equation*}
\sin^2 2 \theta \simeq 1\quad \text{and} \quad 
\Delta m^2 \simeq 3 \times 10^{-3} \evsq \ . 
\end{equation*} 

As is well known, in a two flavor scenario, 
the probability for a muon neutrino with energy $E_\nu$ 
to remain a muon neutrino after propagating the distance $L$
is given by the following expression
\begin{equation}
P_{\mu\mu} = 1-\sin^2 2\theta\sin^2 
\Big(\frac{\Delta m^2 L}{4 E_\nu}\Big) \ .
\label{eq:pmumu}
\end{equation}
Basically, the standard approach to measure the oscillation parameters
is to determine
the oscillation probability in Eq.\ \eqref{eq:pmumu}
in dependence of $E_\nu$. 
At the position of the minimum $\Delta m^2$ can be determined
from the condition $\tfrac{\Delta m^2 L}{4 E_{\nu,{\rm min}}} 
\overset{!}{=} \tfrac{\pi}{2}$ and $\sin^2 2\theta$ from 
$P_{\mu\mu}(E_{\nu,{\rm min}})\overset{!}{=} 1-\sin^2 2\theta$.
The neutrino energy is not directly measurable but
can be reconstructed
from the simple kinematics of quasi-elastic (QE)
scattering events.
Measuring the energy $E_\mu$ and the polar angle $\cos \theta_\mu$ of 
the produced muon allows to reconstruct $E_\nu$
with help of the following relation
(even if the scattered proton is not observed)
\begin{equation}
E_\nu=E_\nu[E_\mu,\cos \theta_\mu] = 
\frac{M E_\mu - \mmu^2/2}{M - E_\mu + |\vec{k}_\mu| \cos\theta_\mu} \ .
\label{eq:E-reconstruction}
\end{equation}
Here $M$ denotes the proton mass, $m_\mu$ the muon mass and 
$\vec{k}_\mu$ is the three-momentum
of the muon in the laboratory system.

However, in practice there are some difficulties.
First of all, the experimental one-ring muon events
($\ormu$) are not pure QE event samples.
About 30$\%$ of the $\ormu$ events are 1-pion production
events with unidentified or absorbed pions.
For the 1-pion events Eq.\ \eqref{eq:E-reconstruction} would
systematically underestimate the true neutrino 
energy \cite{Walter:NuInt02}.
Secondly, the reconstruction of $E_\nu$ gets more complicated
including binding energy $\epsilon_B$ and Fermi motion of the
target nucleons
\begin{eqnarray}
E_\nu &=& E_\nu[E_\mu,\cos \theta_\mu,\vec{p},\epsilon_B]
\\ 
&=& \frac{(E_p+\epsilon_B) E_\mu -
(2 E_p \epsilon_B +\epsilon^2_B+ \mmu^2)/2-\vec{p}\cdot \vec{k}_\mu}
{E_p+\epsilon_B-E_\mu+|\vec{k}_\mu|\cos\theta_\mu-|\vec{p}|\cos\theta_p} \ ,
\nonumber
\label{eq:E-reconstruction1}
\end{eqnarray}
where $\vec{p}$ is the three momentum and 
$E_p = \sqrt{M^2 + \vec{p}^2}$ the energy of the initial nucleon.
Further, $\theta_p$ is the polar angle of the target nucleon w.r.t.\ the 
direction of the incoming neutrino.
Neglecting $\epsilon_B$ and the momentum $\vec{p}\, $ 
Eq.\ \eqref{eq:E-reconstruction} is recovered.
Since the momentum $\vec{p}$ is unknown, $0 \le |\vec{p}| \le p_F$ 
where $p_F$ is the Fermi momentum,
this will lead to an uncertainty of the reconstructed neutrino energy 
at given values $E_\mu$, $\cos\theta_\mu$, and $\epsilon_B$ 
of about -9$\%$ to +6$\%$ for a single event. 

Hence we can see no reliable
way for reconstructing the neutrino energy for 
the $\ormu$ sample on an event by event basis.
On the other hand the muon energy is a directly measurable quantity
for each event. Therefore it seems to us to be a better variable for
testing the spectral distortion phenomenon 
compared to the reconstructed neutrino energy.

In this talk we summarize the basic ideas and the main results 
in \cite{Paschos:2003ej} where we have 
used kinematic considerations to predict
the muon energy spectra of the QE and 1-pion 
resonance production events which constitute the bulk of the
charged-current $\numu$ scattering events in the K2K experiment.
These predictions can be checked with the observed muon energy
spectra from the nearby detector. 
We also present the distortion of these muon
spectra due to $\numu$ oscillation, which one expects to see at
the SK detector. Comparison of the predicted muon spectra with 
those of the observed QE and 1-pion events at the SK detector will be 
very useful in determining the oscillation parameters.

\section{Flux averaged muon energy spectra}

The flux averaged muon energy spectra for QE and 1-pion events 
are given by
\begin{equation}
\big<\frac{{\rm d}\sigma^R}{{\rm d} E_\mu}\big> 
\equiv 
\int f(E_\nu)  \frac{{\rm d}\sigma^R}{{\rm d} E_\mu} {\rm d} E_\nu 
\label{eq:xs}
\end{equation}
where $f(E_\nu)$ is the neutrino flux at K2K for the nearby detector(ND) 
and 'R' denotes QE and $\Delta$ resonance contribution to
1-pion production, which dominates the latter. 
Simple kinematic considerations lead to the following 
approximation for the flux averaged muon energy spectra, both,
for QE and 1-pion production \cite{Paschos:2003ej}
\begin{equation}
\big<\frac{\der \sigma^R}{\der E_\mu}\big>\  
\simeq \sigma_{tot}^R(\overline{E_R})\ f(\overline{E_R})\ , 
\label{eq:app} 
\end{equation}
with
\begin{eqnarray}
\overline{E_R} &=&
E_\mu + \Delta E^R 
=
E_\mu +
\begin{cases}
0.15\ \gev & \text {for \ QE}\\
0.4\ \gev &  \text {for \ 1-pion} \ . \end{cases}
\label{eq:shift}
\end{eqnarray}
Furthermore, it is well known that the total cross sections
for QE and $\Delta$ production tend to constant values for
neutrino energies of about $1\ \gev$ and $1.4\ \gev$, respectively:
$\sigma_{tot}^R[E_\nu] \to N^R$.
Hence, for muon energies larger than about $1.2\ \gev$,
Eq.\ \eqref{eq:app} can be further simplified by replacing 
$\sigma_{tot}^R$ by its constant asymptotic value $N^R$:
\begin{equation}
\big<\frac{\der \sigma^R}{\der E_\mu}\big>\  
\simeq N^R \times f(\overline{E_R})
\quad \text{for}\quad E_\mu \gtrsim 1.2\ \gev \ ,
\label{eq:app2} 
\end{equation}
with
\begin{eqnarray}
N^R &=&
\begin{cases}
4.5\ \fb & \text {for \ QE}\\
5.5\ \fb &  \text {for \ 1-pion} \ . \end{cases}
\label{eq:norm}
\end{eqnarray}
The normalizations correspond to the
average cross-section per nucleon for a 
$H_2O$ target \cite{Paschos:2000be}.

Thus we conclude that at large muon energies 
$E_\mu \gtrsim 1\ \gev$
the flux averaged muon energy cross section 
$\big<\frac{\der \sigma^R}{\der E_\mu}\big>$ is 
directly proportional to the neutrino flux shifted in energy. 
The normalizations $N^R$ and energy shifts $\Delta E^R$ are
predictions of our theoretical calculation \cite{Paschos:2003ej}
which can be verified experimentally with the muon energy
spectra of the QE and 1-pion events observed at the ND.
Furthermore, Eqs.\ \eqref{eq:xs}, \eqref{eq:app}, and \eqref{eq:app2}
also apply to the far detector (FD) if one replaces the flux
at the ND by the flux at the FD which is distorted by the
neutrino oscillation probability $P_{\mu\mu}(E_\nu)$:
\begin{equation}
f(E_\nu) \to \fSK(E_\nu) = f(E_\nu) \times P_{\mu\mu}(E_\nu)\ . 
\label{eq:fdflux}
\end{equation}
Comparing these predictions with the observed muon energy spectra
of the QE and 1-Pion events of the SK detector will test the 
spectral distortion due to $\numu$ oscillation and determine the
oscillation parameters.
Particularly, on the higher energy side of the peak
the relative size of the SK to the ND cross-sections 
provides a direct measure of the spectral distortion and hence the
underlying oscillation parameters:
\begin{eqnarray} 
\frac{\big<\frac{\der \sigma^R}{\der E_\mu}\big>_{{\rm FD}}}
{\big<\frac{\der \sigma^R}{\der E_\mu}\big>_{{\rm ND}}}
\simeq P_{\mu\mu}(\overline{E_R}) 
\quad \text{for}\quad E_\mu \gtrsim 1.2\ \gev \ .
\label{eq:ratio}
\end{eqnarray}

In Sec.\ \ref{sec:numeric} we present exact calculations
of the QE \cite{Paschos:2001np} 
and 1-pion production cross sections.
The dominant contribution to 1-pion production is from 
$\Delta$ resonance production
for which we take the
formalism of Refs.\ \cite{Schreiner:1973mj,Paschos:2002mb,eap:2003}.
For completeness we have also included the $P_{11}(1440)$ and 
$S_{11}(1535)$ resonance contributions for which 
we used the parameterization and the form factors  
of Ref.\ \cite{Fogli:1979cz}.
We estimate the contribution from the still higher resonances 
along with the non-resonant background to be no more than $5-10\%$ of 
the 1-pion production cross section at these low energies 
(see Fig.\ \ref{fig:1}).
Therefore the accuracy of our prediction should be as good 
as that of the K2K experiment.

We compare our exact calculations with the approximation
in Eqs.\ \eqref{eq:app2} and \eqref{eq:ratio} from which
the range of validity can be inferred.
A comparison with the approximation
in Eq.\ \eqref{eq:app} can be found in Ref.\ \cite{Paschos:2003ej}
(see Figs.\ 2 and 3).

\section{Numerical results}
\label{sec:numeric}

\begin{figure}[htb]
\centering
\vspace{-0.8cm}
\hspace{-1.2cm}
\includegraphics[angle=0,width=8.5cm]{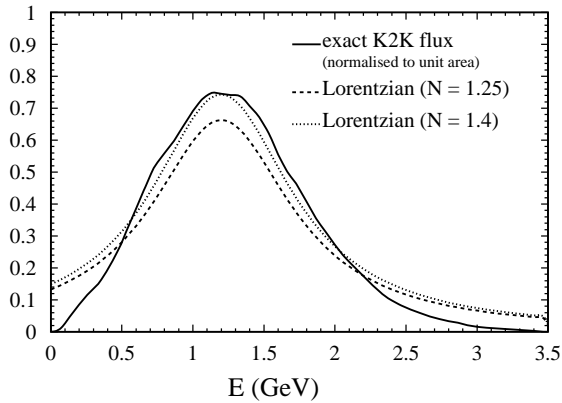}
\vspace*{-1.2cm}
\caption{\sf K2K neutrino energy spectrum. The solid line is the exact
spectrum normalized to unit area. The dashed line shows the 
approximated spectrum by the Lorentzian in Eq.\ (\ref{lorentz}). 
We also show for comparison as dotted line the same Lorentzian, 
but with normalization 1.4 instead of 1.25.}
\label{fig:1}
\end{figure}

In this section we present our numerical results.
Fig. \ref{fig:1} shows the K2K neutrino energy spectrum.
In addition, we show an approximated spectrum 
by a Lorentzian given by
\begin{eqnarray}
f_{\rm{L}}(E_\nu) &=& \frac{N}{\pi}\frac{\Gamma}
{(E_\nu-E_0)^2+\gamma^2}
\label{lorentz}
\nonumber\\
 E_0 &=& 1.2 \ \gev,\ \Gamma = 0.6 \ \gev \ . 
\end{eqnarray}
where $N$ is an appropriate normalization factor.
 
\begin{figure}[htb]
\centering
\vspace*{-0.8cm}
\includegraphics[angle=0,width=8.5cm]{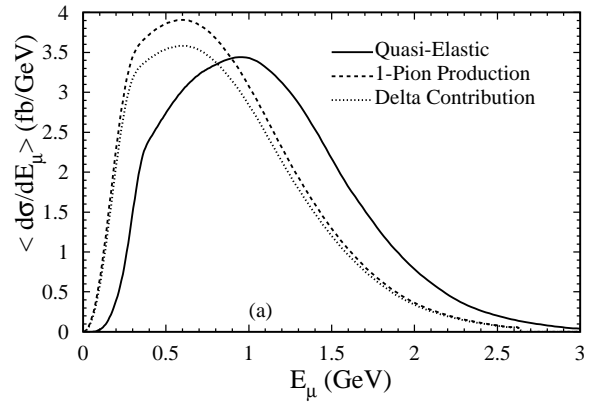}
\includegraphics[angle=0,width=8.5cm]{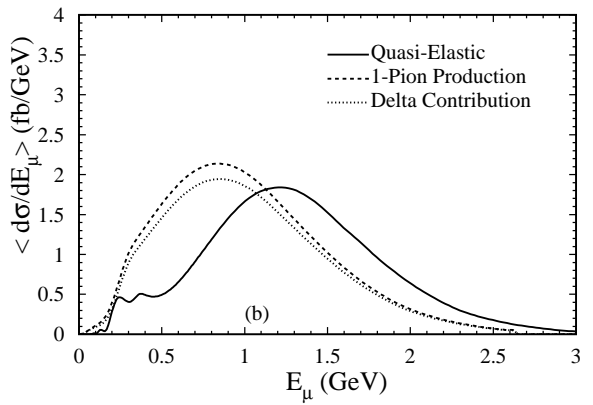}
\vspace*{-1.2cm}
\caption{\sf Exact predictions
 of the muon energy spectra for the (a) Nearby and (b) SK detectors of 
the K2K experiment. The QE (solid line) and 
the 1-pion production (dashed line)  
cross-sections are shown along with the $\Delta$ resonance
contribution (dotted line).}
\label{fig:2}
\end{figure}

Fig.\ \ref{fig:2}a  
shows the predicted muon energy spectra for the QE (solid line) and 1-pion
production (dashed line) processes. 
Clearly one can see that the peak at $E_0= 1.2\ \gev$ is shifted left
by $\Delta E \simeq 0.15\ \gev$ for the QE and
$\Delta E \simeq 0.4-0.5\ \gev$ for the $\Delta$ resonance production 
which dominates
the 1-pion production process. 
The steepness of the muon energy spectra at the low energy  
reflects the threshold rise of $\sigma^R$ and the steep neutrino flux. 
On the other hand the muon energy spectra closely
follow the shape of the neutrino energy spectrum 
on the right side of the peak.
The predicted exact muon energy spectra of Fig.\ \ref{fig:2}a agree
reasonably well with the
corresponding spectra of the K2K ND\ \cite{Ahn:2002up} for both the 
QE and the non-QE sample. 
In particular one can compare the predicted QE spectrum with their 
simulated QE spectrum shown in Fig.\ 1 of Ref.\ \cite{Ahn:2002up}. 
Their Figs.\ 1a and c
show separately the QE muon momentum distribution
for the 1-Ring muon ($\ormu$) sample of the
1KT and the QE enhanced sample of the FGD respectively. 
The two play complementary
roles in covering the complete muon energy range, 
as the 1KT and the FGD have
high efficiencies at $E_\mu < 1\ \gev$ and $E_\mu \gtrsim 1\ \gev$, 
respectively \cite{Ahn:2002up}. 
One can not compare our predicted muon energy spectra with these figures
quantitatively without folding in these efficiency factors, which
are not available to us.
But
there is good qualitative agreement between the predicted QE spectrum 
of our Fig.\ \ref{fig:2}a with their Fig.\ 1c at $E_\mu \gtrsim 1\ \gev$ 
and Fig.\ 1a at
$E_\mu < 1\ \gev$. While the former shows the position of the peak and 
the shape of the spectrum to the right, the latter shows the broadening of
spectrum down to $E_\mu \simeq 0.4\ \gev$. 
Similarly one
sees good agreement between the predicted muon energy spectrum of our 
Fig.\ \ref{fig:2}a for
1-pion events with the non-QE spectra of their Fig.\ 1c,d at 
$E_\mu \gtrsim 1\ \gev$
and Fig.\ 1a at $E_\mu < 1\ \gev$. 
Thus one has a simple and robust prediction
for the shape of the muon energy 
spectrum in terms of the neutrino spectrum not only
for the QE events but also for the 1-pion production events, 
which dominate the inelastic events.

Fig.\ \ref{fig:2}b shows the corresponding muon energy spectra of the
QE and 1-pion events for the SK detector, predicted by 
Eqs.\ \eqref{eq:xs} and \eqref{eq:fdflux}.
One can clearly see the distortion of the muon energy spectrum due the
$\numu$ oscillation.
They should be compared with the observed muon energy spectra of the
1-Ring and 2-Ring muon events at the SK detector, after taking into account
the pion detection efficiency. We hope such a comparison 
will be done by the K2K collaboration.

For the estimation of the pion detection 
efficiency it is necessary to consider
nuclear effects, i.e., Pauli blocking, nuclear absorption
and charge exchange of the produced pions 
which can rescatter several times in the nucleus. 
Therefore we have included these effects 
following the prescription of 
Refs.\ \cite{Adler:1974qu,Paschos:2000be,Schienbein:2003sm}.
Since the dominant contribution to 1-pion production processes 
comes from $\Delta$ resonance production on oxygen, 
we have evaluated the effects of nuclear absorption and rescattering
on the produced pions for this case.
The relevant charged current subprocesses are
$\nu p \to \mu^- p \pi^+$, $\nu n \to \mu^- n \pi^+$ and 
$\nu n \to \mu^- p \pi^0$ with relative cross-sections $9:1:2$ 
for the dominant contribution from $\Delta$ resonance.

\begin{figure}[htb]
\centering
\vspace*{-0.3cm}
\includegraphics[angle=0,width=8.5cm]{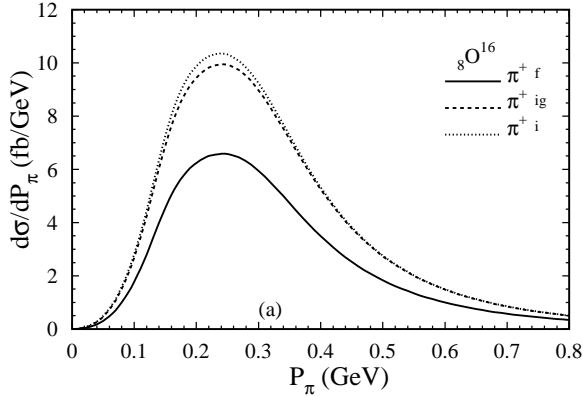}
\includegraphics[angle=0,width=8.5cm]{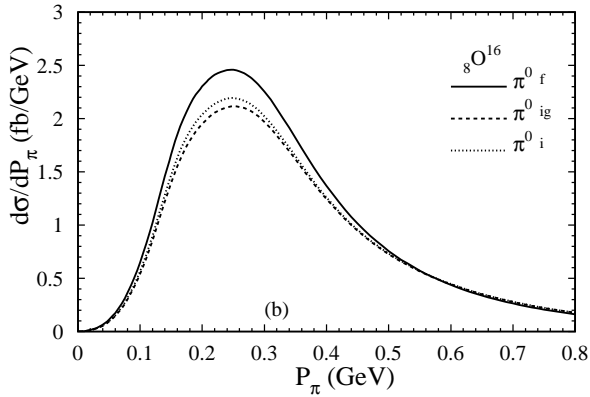}
\vspace*{-1.2cm}
\caption{\sf The momentum distribution of the decay pion for 
charged current 
resonance production in oxygen with (solid line) 
and without nuclear correction (dotted line). The dashed line 
takes only into account the effect of Pauli blocking.}
\label{fig:5}
\end{figure}

Fig.\ \ref{fig:5} shows the effects of nuclear corrections on the 
produced $\pi^+$ and $\pi^0$ spectra from these processes 
for the nearby detector averaged over the neutrino spectrum. 
The results are very similar for the SK detector.
Nuclear rescattering effects result in enhancing the $\pi^0$ events
at the cost of the dominant $\pi^+$ component. But taken together we see
a nearly $20 \%$ drop in the rate of 1-pion events due to nuclear absorption
of the produced pion.
Moreover about $10 \%$ of the remaining events corresponds 
to the pion momentum
being less than the Cerenkov threshold of $100\ \mev$. Therefore one
expects about $70\%$ of the $\Delta$ events to give a detectable
pion ring at the SK detector while the remaining $30\%$ appears like
a QE event.
Adding the latter to the $35 \%$ of genuine QE events would imply
that about $50 \%$ of the CC events will appear QE-like at the SK detector.
Alternatively the observed muon energy spectrum
of the sum of 1-Ring and 2-Ring muon events could be compared with
the predicted spectrum of the sum of QE and 1-Pion events.
Although a part of the 2-Ring events may come from multi-pion
production, the resulting error may be small since multi-pion events at 
the ND constitute only $\sim 15 \%$ of CC events.

\begin{figure}[htb]
\centering
\vspace*{-0.9cm}
\includegraphics[angle=0,width=8.5cm]{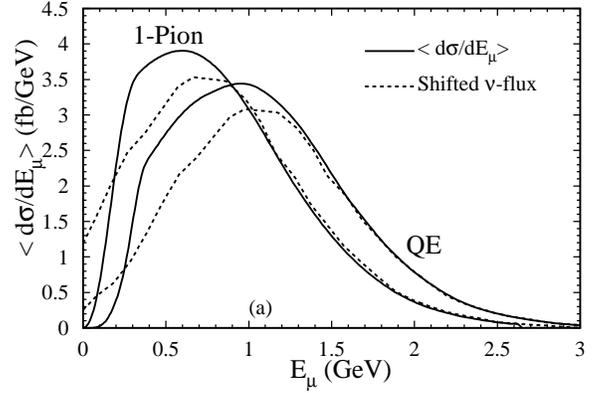}
\includegraphics[angle=0,width=8.5cm]{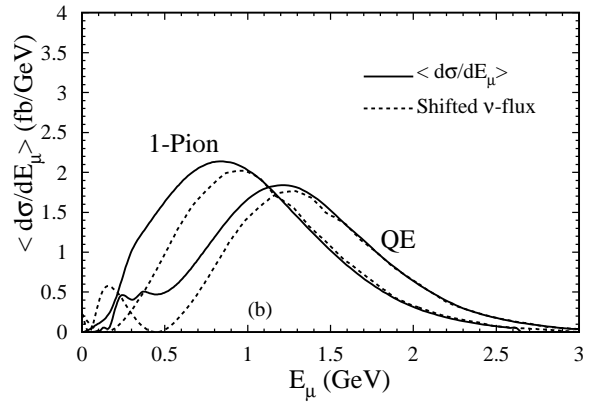}
\vspace*{-1.2cm}
\caption{\sf The predicted exact 
muon energy spectra (solid lines) and  
the approximation (dashed lines) according to
Eq.\ \protect\eqref{eq:app2}
for QE and 1-pion production for the (a) Nearby detector and 
(b) SK detector of the K2K experiment.}
\label{fig:3}
\end{figure}

Next, we turn in Fig.\ \ref{fig:3} to a comparison of 
the exact calculation for the muon energy spectra (solid lines) 
with the approximation (dashed lines) made on the
right hand side of Eq.\ \eqref{eq:app2}.
Fig.\ \ref{fig:3}a shows the results for the ND
and Fig.\ \ref{fig:3}b for the FD obtained with the
neutrino flux specified in Eq.\ \eqref{eq:fdflux}.
One can see that the approximation is in perfect agreement
with the exact calculation for muon energies 
$E_\mu \ge 1.2\ \gev$ for single pion production
and $E_\mu \ge 1\ \gev$ for QE scattering.
Hence, in this region the muon energy spectra
are directly proportional to the neutrino flux shifted in energy.

\begin{figure*}[htb]
\centering
\vspace*{-0.9cm}
\includegraphics[angle=0,width=8.5cm]{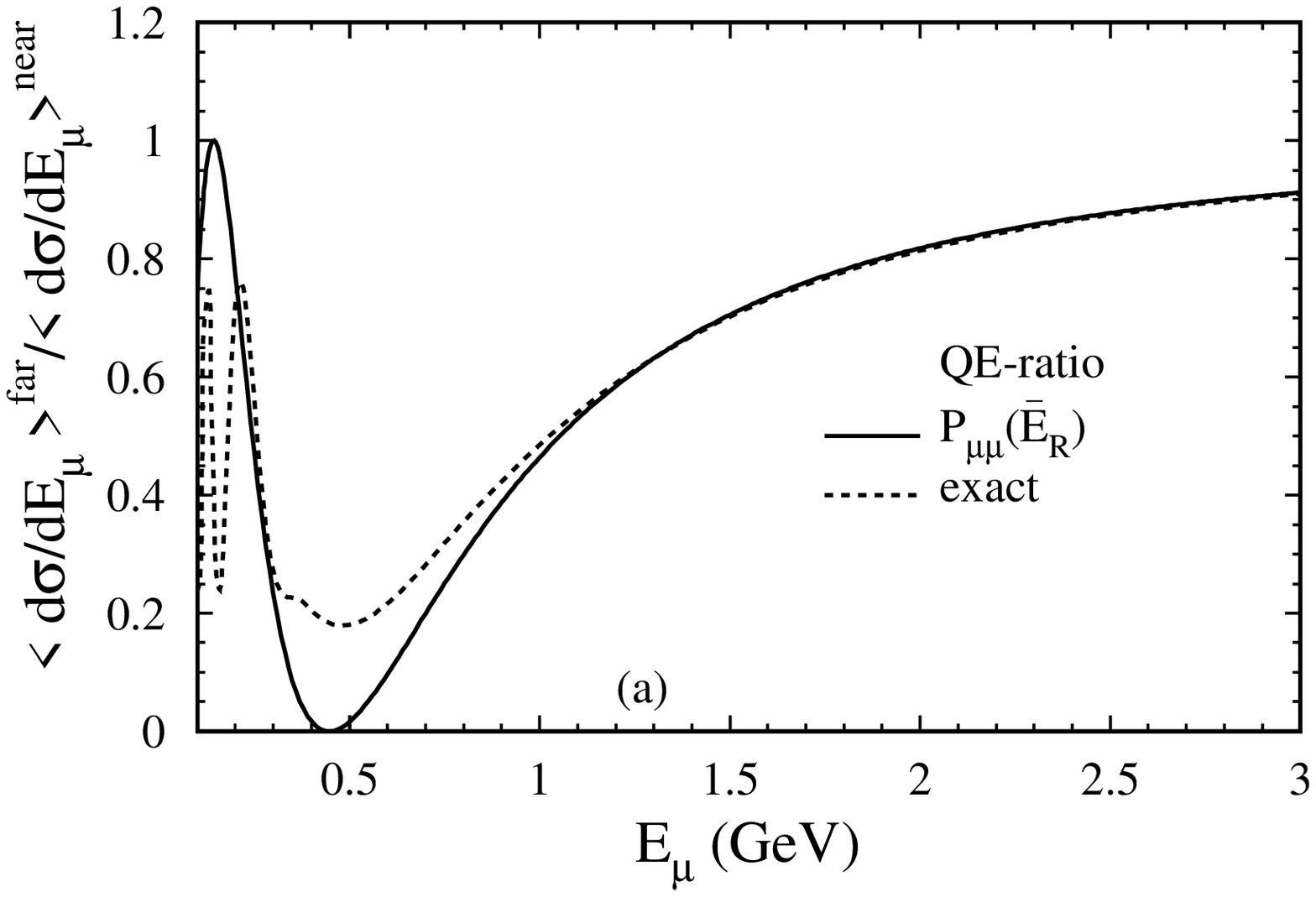}
\includegraphics[angle=0,width=8.5cm]{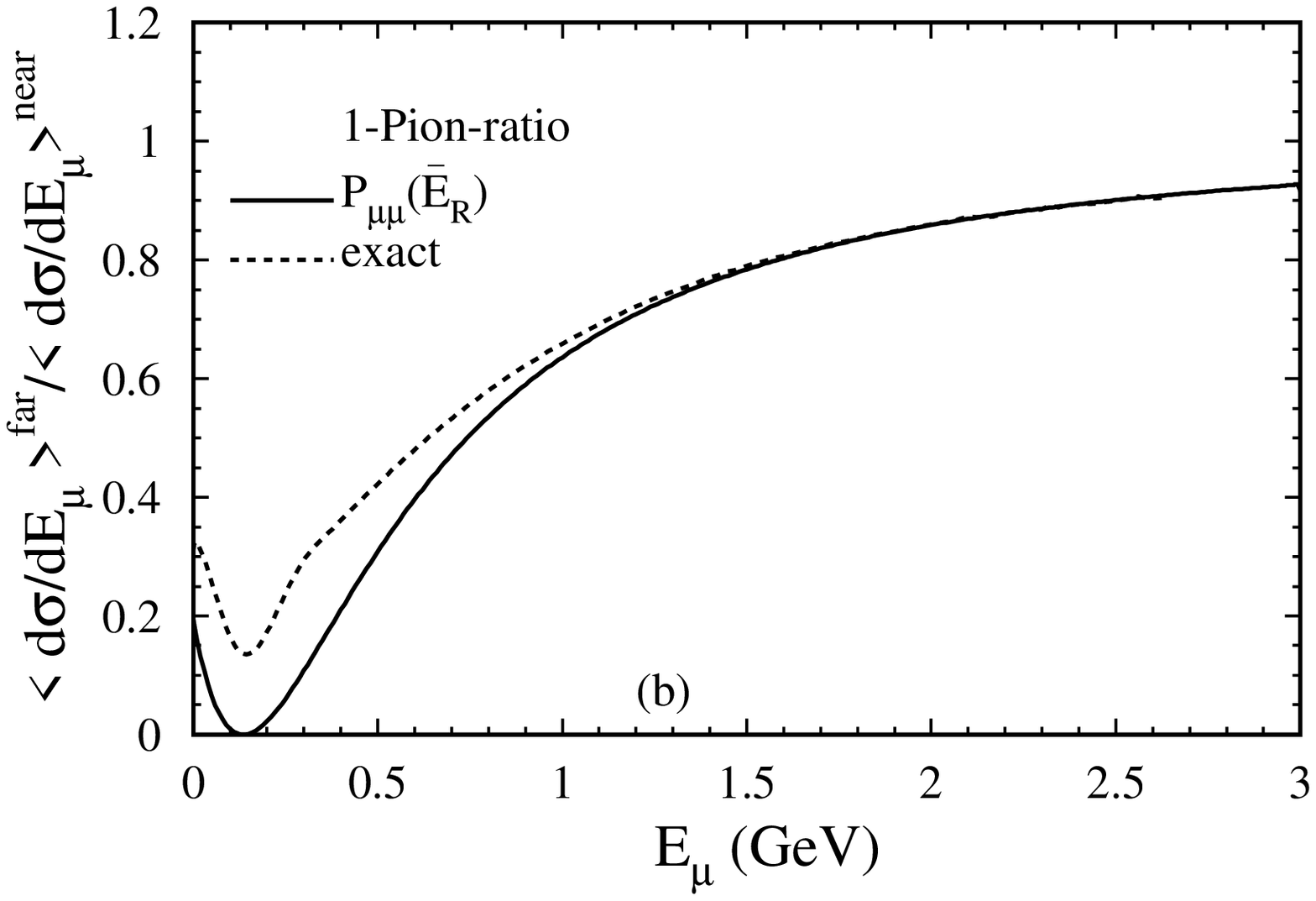}
\vspace*{-1.2cm}
\caption{\sf The predicted 
ratio of the far and nearby detector of the K2K experiment 
for (a) QE and (b) 1-pion production. 
The solid line is the neutrino oscillation probability.
The dashed line denotes the predicted exact 
ratio of the far and near detector of the K2K experiment.}
\label{fig:4}
\end{figure*}

Finally, in Figs.\ \ref{fig:4}a and \ref{fig:4}b  
we show the far-near-ratio of the muon energy spectra 
for QE and 1-pion production, respectively.
In this case, the dashed lines depict the exact result and
the solid lines the oscillation probabilities 
$P_{\mu\mu}$ given in Eq.\ \eqref{eq:pmumu}
which have been
evaluated at the shifted energies $\overline{E_R}$ given 
in Eq.\ \eqref{eq:shift}.
Again, at $E_\mu \ge 1\ \gev$ (QE) and $E_\mu \ge 1.2\ \gev$
(1-pion production) the approximation in Eqs.\ \eqref{eq:app2}
and \eqref{eq:ratio} works well and a measurement of the 
far-near-ratio of such pure QE or 1-pion event samples 
in this kinematic region would give direct access to the
oscillation probability.
However, as has been discussed above,
the observable
1-Ring and 2-Ring muon events are superpositions of
QE, 1-pion, and multi-pion events making the analysis
more complicated.

\section{Conclusions}
The muon energy spectra of QE and 1-pion events 
provide a complementary approach 
to experimental extractions of the atmospheric 
muon oscillation parameters at K2K. 
The results are based on quite general kinematic considerations
and will also
be applicable to other
future long baseline experiments
like J2K, MINOS and the CERN-Gran Sasso experiments which plan to
use low energy $\numu$ beams \cite{Itow:NuInt01,Lipari:NuInt01}.
Therefore it will be very useful to extend this analysis for the
beam energy spectra and the target nuclei of these
experiments.

\noindent{\large\bf{Acknowledgment}}\\
J.-Y. Yu wishes to thank the organizers of the ICFP03 in Seoul
for the kind invitation and financial support.

\end{document}